\documentclass[]{spie}  %>>> use for US letter paper
\usepackage{amsmath,amsfonts,amssymb}
\usepackage{graphicx}
\usepackage{url}
\usepackage[colorlinks=true, allcolors=blue]{hyperref}

\title{A fast, wide-field, and real-time imaging prototype for large aperture arrays}

\author[a]{Nithyanandan Thyagarajan}
\author[b]{Jishnu N. Thekkeppattu}
\author[b]{David Humphrey}
\affil[a]{Space \& Astronomy, Commonwealth Scientific and Industrial Research Organisation (CSIRO), P. O. Box 1130, Bentley, WA 6102, Australia}
\affil[b]{Space \& Astronomy, Commonwealth Scientific and Industrial Research Organisation (CSIRO), P. O. Box 76, Epping, NSW 1710, Australia}

\authorinfo{Further author information: (Send correspondence to N.T.)\\N.T.: E-mail: Nithyanandan.Thyagarajan@csiro.au, Telephone: +61 8 6436 8626}

\begin{document} 
\maketitle

\begin{abstract}
Real-time processing on sub-millisecond timescales is essential for detecting fast astrophysical transients such as fast radio bursts (FRBs) and prompt electromagnetic counterparts to gravitational-wave mergers. Immediate localisation enables rapid multi-wavelength follow-up and maximises scientific return. As a result, real-time capability has become a key requirement for modern wide-field aperture arrays, which are increasingly being deployed at scales of thousands to tens of thousands of antenna elements.
However, traditional correlator architectures scale as $\mathcal{O}(N^2)$, creating significant computational, bandwidth, and power challenges for large-$N$ arrays. These constraints often limit field of view, angular resolution, or observing duty cycle, ultimately reducing transient discovery potential.

Direct imaging of the full field of view using E-field Parallel Imaging ``Correlator'' (EPIC) offers a promising alternative by avoiding the explicit formation of pairwise correlations. EPIC can achieve $\mathcal{O}(N \log N)$ computational scaling for dense aperture arrays and has been demonstrated on the Long Wavelength Array (LWA) station in Sevilleta (USA) using Graphics Processing Unit (GPU). However, GPU implementations can be limited by memory bandwidth in low-bitwidth gridding and Fast Fourier Transform (FFT) workloads. We, therefore, are developing a prototype implementation of EPIC using Field Programmable Gate Arrays (FPGA) that exploits highly parallel FFT pipelines to improve computational efficiency and reduce power consumption. Using preliminary results, we evaluate its performance particularly from the viewpoint of scalability, and discuss its suitability for next-generation instruments including SKA-Low and mid-frequency aperture arrays.

\end{abstract}


\section{INTRODUCTION}
\label{sec:intro}  % \label{} allows reference to this section

There is a rapid increase in discoveries of astrophysical time-domain phenomena at radio wavelengths. These span a range of emission mechanisms such as in pulsars \cite{Keane2013}, fast radio bursts (FRB) \cite{Lorimer+2007,Thornton+2013}, magnetars and their potential link to FRBs \cite{Bochenek+2020}, rotating radio transients (RRAT) \cite{Mclaughlin+2006}, long period transients (LPT) \cite{Rea+2026}, gamma-ray burst (GRB) afterglows, Jovian and solar bursts, flare stars, cataclysmic variables (CVs), exoplanet and stellar outbursts \cite{Zhang+2023}, X-ray binaries, novae, supernovae, active galactic nuclei (AGNs), blazars, tidal disruption events, and counterparts to gravitational wave (GW) events. These phenomena also exhibit a tremendous variety of activity timescales ranging from nanoseconds to days, across wide frequency ranges and polarisation states \cite{Pietka+2015,Chandra+2016,Nimmo+2022}. The prospects of discovering new phenomena are ever increasing. 

The time-domain discovery figure of merit is described by several factors \cite{Murphy+Kaplan2026}. The likelihood of discovery of transient phenomena is increased by the instantaneous field of view (inversely proportional to array element size), the number of epochs on the timescale of interest, and sensitivity which in turn depends on the collecting area, bandwidth and accumulation timescale. Driven by these factors, modern aperture arrays that include time-domain studies in their key science aim to achieve sensitivity and wide field of view through a large number of small-sized elements. Examples include the Bustling Universe Radio Survey Telescope in Taiwan (BURSTT) \cite{BURSTT}, the Deep Synoptic Array (DSA) \cite{DSA-2000}, \texttt{SKA-low} \cite{Dewdney+2009,SKA1+2019}, Murchison Widefield Array (MWA) \cite{Tingay+2013}, the Low Frequency Array (LOFAR) \cite{vanHaarlem+2013}, the swarm of Long Wavelength Arrays (LWA Swarm) \cite{Dowell+2018}, the Hydrogen Intensity and Real-time Analysis Experiment (HIRAX) \cite{HIRAX+2022}. 

Although large aperture arrays can be flexible in their architecture and operations \cite{Thyagarajan2025}, their demands are high in computational performance and data throughput \cite{Thyagarajan+2017,Thyagarajan+2019}. The computational and bandwidth load on traditional correlators that output visibilities scale as $\mathcal{O}(N_\textrm{e}^2)$, where, $N_\textrm{e}$ is the number of array elements in the aperture array \cite{Thyagarajan2025}. Modern arrays like the \texttt{DSA-2000} are planning to employ a correlator architecture but will convert the visibilities to images and save on the output data volume by not requiring to store visibilities \cite{DSA-2000}. Traditional beamformers can perform coherent sums of complex voltages in several directions simultaneously. To cover the full field of view, the computational cost will scale with $N_\textrm{e}$ and number of beams required to fill the field of view. If these costs become unsustainable, typically correlator-based architectures will have to compromise on spatial and/or temporal resolution. Similarly, beamformers will be forced to compromise on the field of view and/or spatial resolution \cite{Price2024}. Such compromises could severely curtail their potential to discover transient phenomena. 

Direct imaging with spatial Fast Fourier Transform (FFT) architectures can salvage the situation for large aperture arrays by exploiting the FFT's computational efficiency while avoiding the need for spatial correlations. Initially, only array layouts with elements constrained to be on a regular grid in order to be able to apply the FFT were considered \cite{Daishido+1991,Otobe+1994,Foster+2014}, which can be disadvantageous for several imaging applications. However, a generalised direct imaging architecture that can be applied to arbitrary spatial layouts while still offering FFT's efficiency, called the E-field Parallel Imaging ``Correlator'' (EPIC) \cite{Thyagarajan+2017} was formulated. EPIC's solution involves gridding of voltages, spatial FFT of the gridded voltages, and pixelwise ``squaring'' (or outer product of the polarisation states) to obtain polarimetric images filling the entire field of view with high temporal resolution. EPIC offers most benefits over traditional correlator and beamformer architectures when the aperture array consists of a larger number of elements packed densely \cite{Thyagarajan+2017,Thyagarajan+2019,Thyagarajan2025} and when imaging the full field of view is required, where it approaches an efficiency $\sim N_\textrm{e}\log N_\textrm{e}$. 

EPIC has been successfully deployed on the Long Wavelength Array (LWA) station in Sevilleta (New Mexico, USA) on a GPU platform \cite{Kent+2019,Kent+2020,Krishnan+2023,Reddy+2024}. The advantage that the LWA-Sevilleta gained from this EPIC deployment is exemplified by the improvement in output cadence of imaging from $\simeq 5$~s in the case of a software correlator to $\simeq 80$~ms using a GPU-based EPIC \cite{Reddy+2024} with a processing bandwidth of 19.2~MHz. EPIC is offered as a commensal backend system to the users of LWA-Sevilleta \cite{Krishnan+2023}. Despite this enormous improvement over a conventional correlator-based architecture, it is estimated that the scalability of EPIC on GPUs is limited even on advanced systems like the \texttt{H100} because of the shared memory constraints \cite{LWA-EPIC-memo9}. 

This motivates our exploration of deploying EPIC using a Field Programmable Gate Array (FPGA) platform. A FPGA-based implementation of EPIC promises to be scalable to larger aperture arrays and image sizes due to their efficiency in implementing low bitwidth FFT operations and high bandwidth memory (HBM) available on commercial FPGA cards. 

The paper is organised as follows. Section~\ref{sec:EPIC} introduces the concept of EPIC. Section~\ref{sec:FPGA} describes the FPGA prototype that is under development. Current status and results are provided in section~\ref{sec:results}. The prospects of scalability of the FPGA-based implementation to future aperture arrays are highlighted in section~\ref{sec:future}. We summarise the work in section~\ref{sec:summary}.

\section{E-field Parallel Imaging ``Correlator''}\label{sec:EPIC}

\begin{figure} [ht]
   \begin{center}
   \begin{tabular}{c} %% tabular useful for creating an array of images 
   \includegraphics[height=8cm]{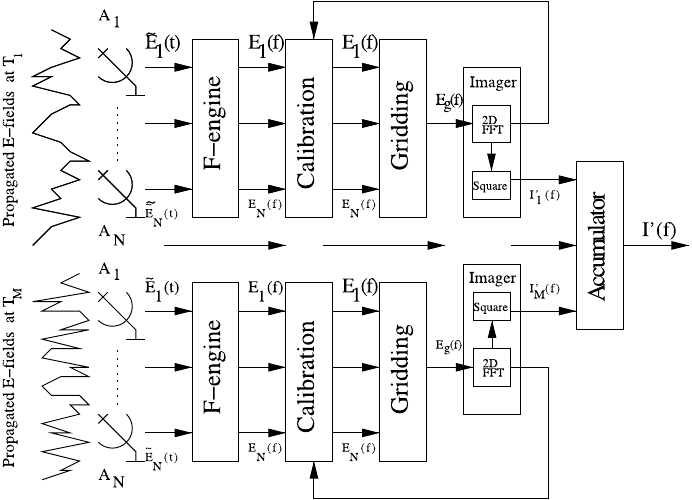}
   \end{tabular}
   \end{center}
   \caption[EPIC architecture] 
%>>>> use \label inside caption to get Fig. number with \ref{}
   { \label{fig:EPIC} Architecture of EPIC. Equations~\ref{eqn:intra-station-pol-hol-img-epic} and \ref{eqn:intra-station-opt-pol-img-outprod} are applied in blocks that appear after channelisation (F-engine) and calibration. (Reproduced from the original EPIC paper \cite{Thyagarajan+2017}.)}
\end{figure}

EPIC consists of a gridding operation on the calibrated voltages, $E_e^p$, of polarisation, $p$, on array element, $e$, with a complex gridding kernal that transforms it into a grid voltage of output polarisation, $\alpha$, which is then followed by a spatial FFT expressed as \cite{Thyagarajan+2017,Thyagarajan2025},
\begin{align}
    \mathcal{E}^\alpha(\hat{\boldsymbol{s}}_k) &= \sum_j \delta^2 \boldsymbol{r}_j \, e^{i\frac{2\pi}{\lambda} \hat{\boldsymbol{s}}_k\cdot\boldsymbol{r}_j} \left(\sum_e \sum_p W_e^{\alpha p*}(\boldsymbol{r}_j-\boldsymbol{r}_e) \, E_e^p \right) \, . \label{eqn:intra-station-pol-hol-img-epic}
\end{align}
The expression within the parenthesis represents the gridding operation of each calibrated electric field measurement at an arbitrary location, $\boldsymbol{r}_e$, of element $e$ onto a common grid at locations, $\boldsymbol{r}_j$, using a gridding kernel, $W_e^{\alpha p*}(\boldsymbol{r})$ associated with that element. The outermost summation denotes the Fourier transform of the weighted and gridded electric fields implemented through FFT. Application of the FFT will have the effect of simultaneously beamforming over the entire field of view, $\Omega_\textrm{e}$. Finally, the polarised intensities are obtained by pixelwise ``squaring'' (outer product of polarisation states that reduces to squaring when $\alpha=\beta$) and temporal averaging,
% The polarised intensity in the beamformed pixel is then obtained by
\begin{align}
    \mathcal{I}^{\alpha\beta}(\hat{\boldsymbol{s}}_k) &= \left\langle \mathcal{E}^\alpha(\hat{\boldsymbol{s}}_k) \,  \mathcal{E}^{\beta *}(\hat{\boldsymbol{s}}_k) \right\rangle \, , \label{eqn:intra-station-opt-pol-img-outprod}
\end{align}
where, angular brackets denote a temporal averaging across an interval of $t_\textrm{acc}$. Superscript, $\alpha\beta$, denotes the four pairwise combinations of polarisation states of the intensity. Figure~\ref{fig:EPIC} shows the architecture of EPIC as various operational blocks.

\section{FPGA prototypes}\label{sec:FPGA}

We use commercially available \texttt{AMD Alveo V80} FPGA accelerator card to implement EPIC. The \texttt{AMD Alveo V80} card has 4x200GbE interfaces, 32~GB of HBM2e memory, 2.6M lookup tables (LUT), 10848 DSP elements, 673~MB of on-chip memory. This card is identical to that the SKA-Low correlator and beamformer (CBF) uses \cite{Hampson+2025}. Channelised and calibrated complex voltages for the aperture array elements are loaded into the HBM, from which the EPIC core reads the data and performs processing as outlined in section~\ref{sec:EPIC}. The FFT core is designed to be flexible to suit different scaling schemes.

\begin{figure} [ht]
   \begin{center}
   \begin{tabular}{cc} %% tabular useful for creating an array of images 
   \includegraphics[height=4.5cm]{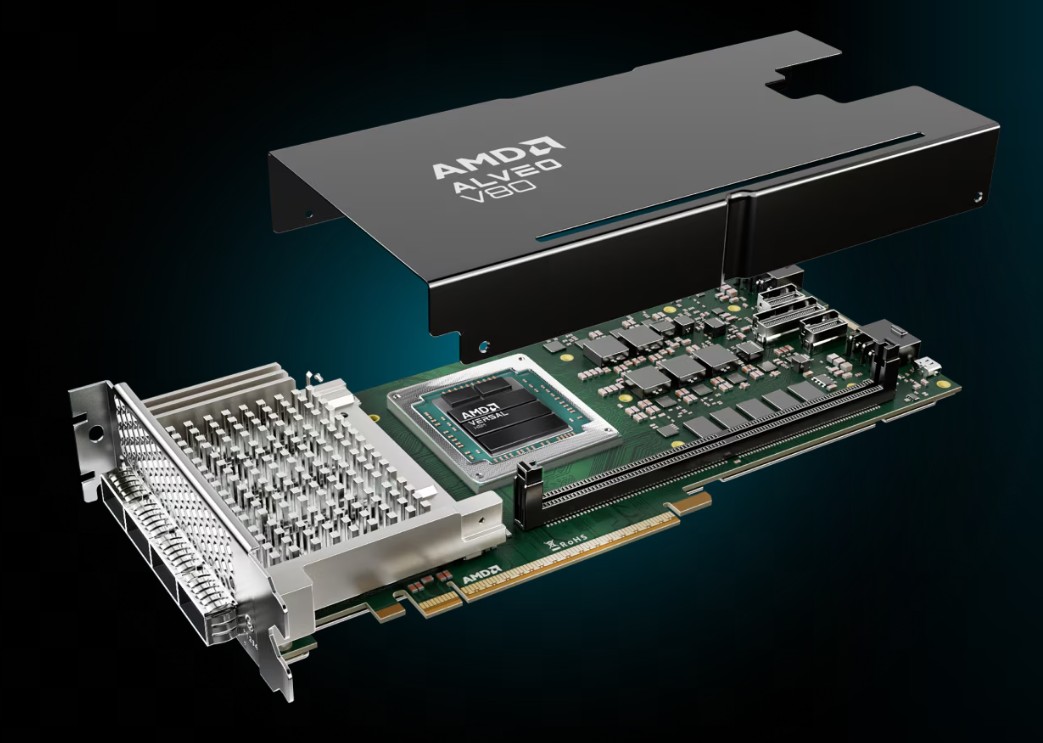} & 
   \includegraphics[height=5cm]{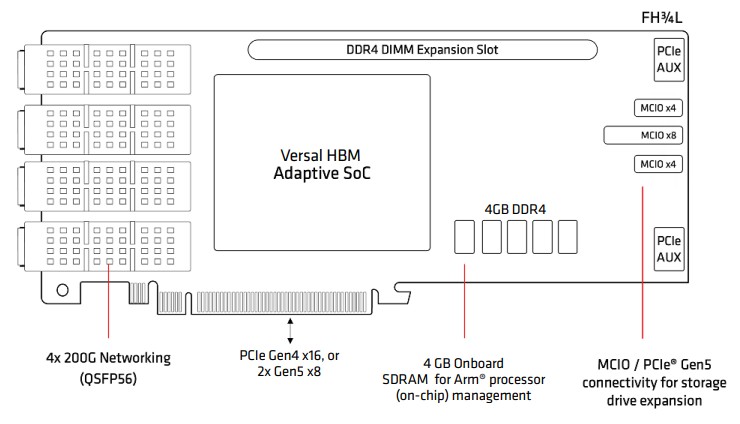}
   \end{tabular}
   \end{center}
   \caption[\texttt{AMD Alveo V80} card] 
%>>>> use \label inside caption to get Fig. number with \ref{}
   { \label{fig:V80} Features of an \texttt{AMD Alveo V80} card. (Source: AMD webpage)}
\end{figure} 

We are developing two prototypes to demonstrate the operability of FPGA-based EPIC on different scales. 
\begin{enumerate}
    \item Prototype to form 64$\times$64 images from a 256-element dual-polarisation aperture array with an all-sky field of view, a bandwidth of $\sim 50$~MHz, and an output frame rate of $\gtrsim 1000$ frames per second ($\lesssim 1$~ms output cadence) on a single \texttt{AMD Alveo V80} card. The layout chosen for this aperture array is based on the Aperture Array Verification System 2.0 (AAVS2) \cite{Macario+2022}, and is typical of an SKA-low station. 
    \item Prototype to form 512$\times$512 all-sky field of view images from an aperture array consisting of $\sim 20,000$ dual polarisation elements. This would be comparable to simultaneously processing elements in as many as $\simeq 20-30$ core SKA-low stations, thus being able to support a tile processing module's (TPM) coarse channel per \texttt{AMD Alveo V80} card. 
    With a coarse channel bandwidth of $\simeq 780$~kHz, the aim is to achieve an imaging cadence of $\lesssim 1$~ms, but the actual frame rate will be subject to I/O constraints.
\end{enumerate}

\begin{figure} [ht]
   \begin{center}
   \begin{tabular}{cc} %% tabular useful for creating an array of images 
   \includegraphics[height=7cm]{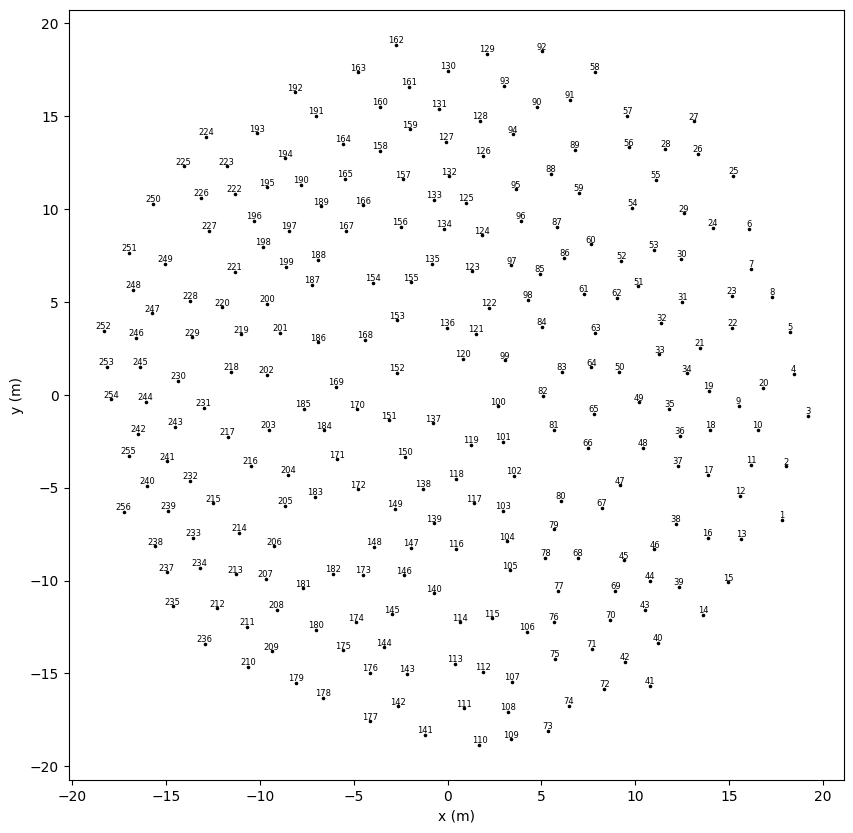} & 
    \includegraphics[height=7cm]{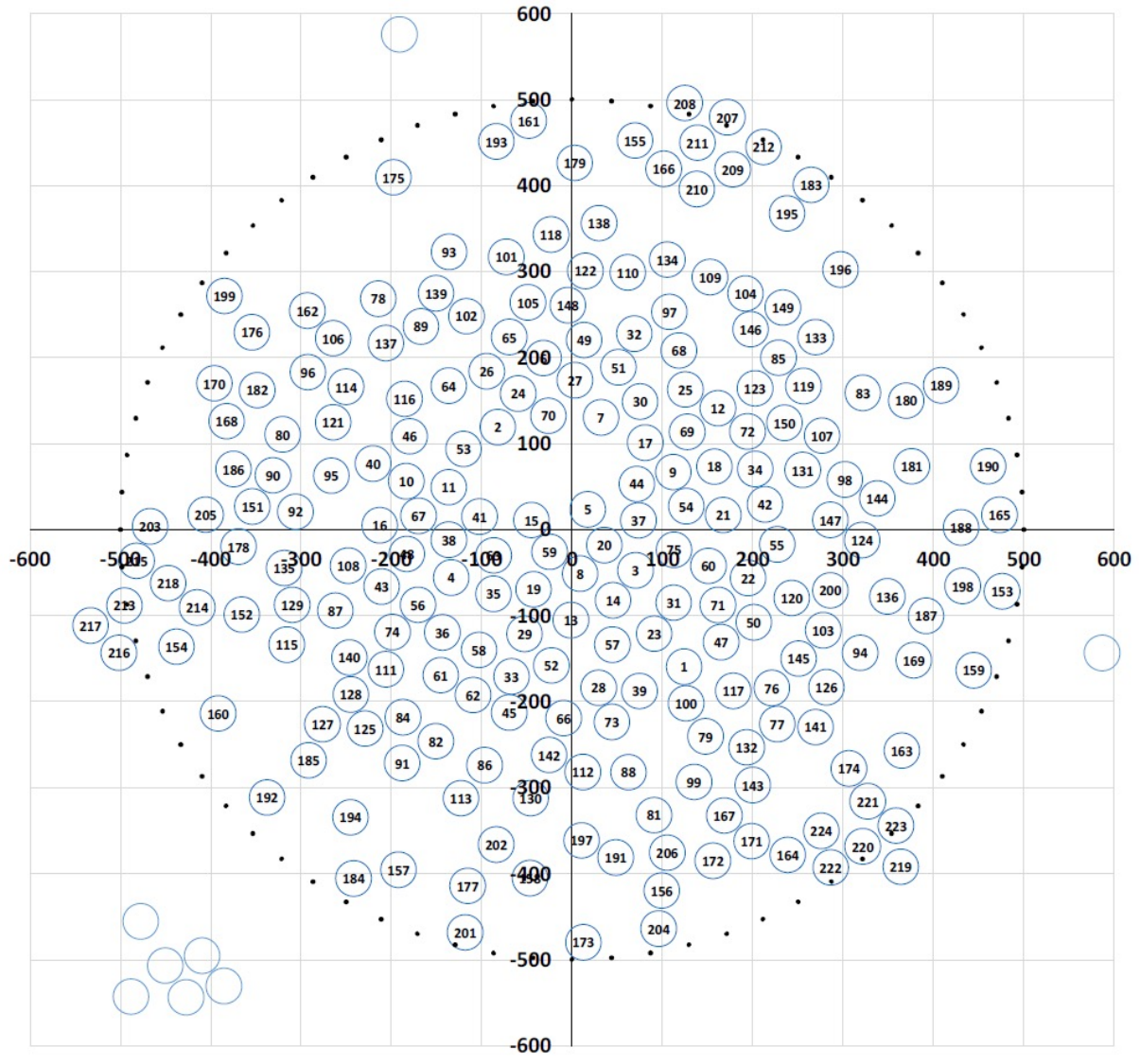}
   \end{tabular}
   \end{center}
   \caption[Aperture array layouts] 
%>>>> use \label inside caption to get Fig. number with \ref{}
   { \label{fig:array-layouts} \textit{(Left)} Antenna layout of the AAVS2 station. \textit{(Right)} Station layout in the core of SKA-low.}
\end{figure} 

% There are scaling and rounding steps between the FFT stages to prevent saturation effects. 

\section{Results}\label{sec:results}

We currently do not have operational aperture arrays to provide real-time streaming data. So, we created synthetic data using the AAVS2 layout, which is a full-scale engineering prototype station for the \texttt{SKA-low} radio telescope located at Inyarrimanha Ilgari Bundara -- CSIRO Murchison Radio-astronomy Observatory. The operating frequency of AAVS2 is taken to be 159~MHz. We used three persistent sources of emission and injected a transient source in one of the frames each with their intrinsic randomness. No additional thermal noise was added. 

The synthetic complex dual polarisation voltage data from 256 elements were digitised using 8 bits each for real and imaginary parts. The synthetic voltages were gridded at finer than half-wavelength grid spacing (to get a power of two for the grid size and to avoid spatial aliasing) using a 2$\times$2 complex gridding kernel
% represented using $8+8$ bits, 
which were then Fourier transformed with a 
% 64$\times$64 
two-dimensional FFT kernel.
% with $8+8$-bit complex twiddle factors. 
% The choice of bitwidth was determined by experimentation that struck a balance between signal-to-noise ratio and the resources on the \texttt{AMD Alveo V80} card. 
With optimisations for low-bitwidth operations, EPIC is able to sustain polarisation capabilities and generate pseudo-Stokes maps at the output.

\begin{figure} [ht]
   \begin{center}
   \begin{tabular}{cl} %% tabular useful for creating an array of images 
   \includegraphics[height=6.8cm]{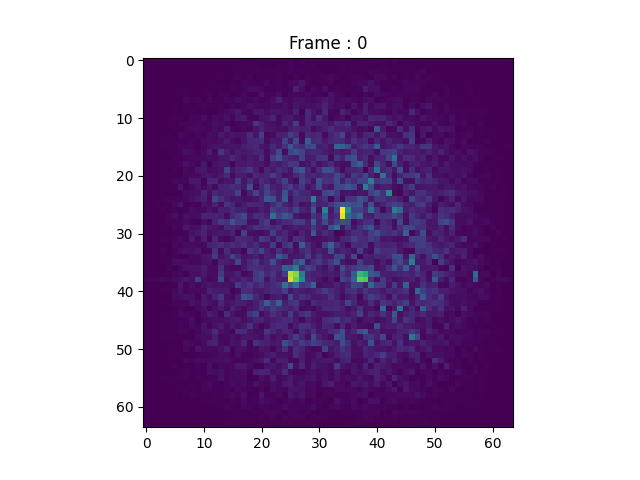} & 
    \includegraphics[height=6.8cm]{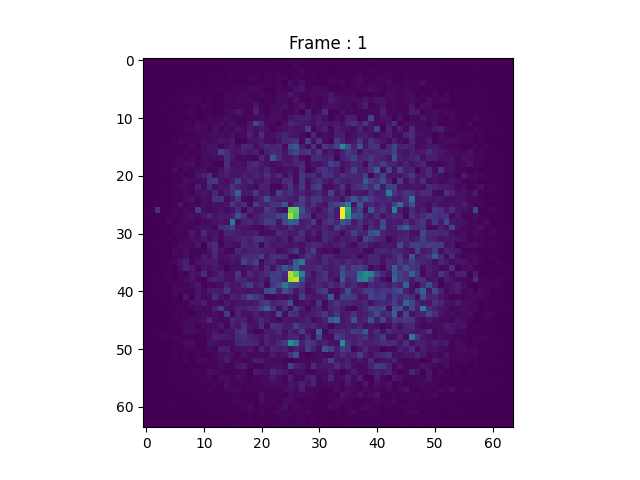} \\ 
   \includegraphics[height=6.8cm]{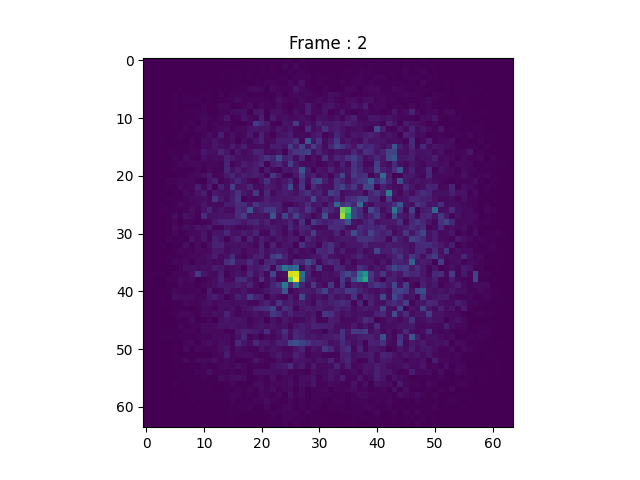} & 
    \includegraphics[height=6.8cm]{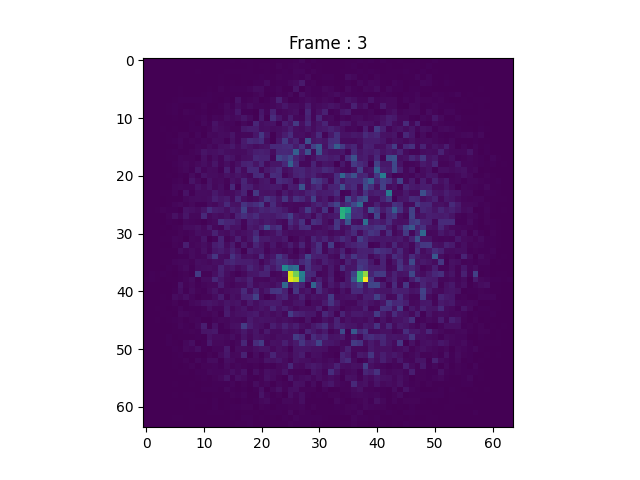} 
   \end{tabular}
   \end{center}
   \caption[Image frames] 
%>>>> use \label inside caption to get Fig. number with \ref{}
   { \label{fig:images} Four 64$\times$64 image frames from synthetic data for an AAVS2 layout (see figure~\ref{fig:array-layouts}) processed through the AMD Vivado simulator. All four pseudo Stokes images are produced, but only one of the copolar images is shown. Each frame here is an accumulation of four individual frames at 1.28~$\mu$s cadence. Thus, the net cadence corresponds to 5.12~$\mu$s. The presence of an injected transient source of emission (top left in the frame 1) is noted among three persistent sources. The synthetic data contains intrinsic randomness from the sources themselves but no additional thermal noise.}
\end{figure}

We assumed a channel width of 780~kHz, and thus every processed frame corresponds to an interval of 1.28~$\mu$s. Each output frame is formed from accumulating four processed frames. Thus, the cadence of each output image frame is 5.12~$\mu$s. The images so formed with an AMD Vivado simulator are shown in Figure~\ref{fig:images}, illustrating that the persistent sources and the transient source are recovered at the correct location on the sky. This demonstrates that our timing and ordering of the data, and FFT implementations are accurate.

\begin{figure} [ht]
   \begin{center}
   \begin{tabular}{c} %% tabular useful for creating an array of images 
   \includegraphics[height=7cm]{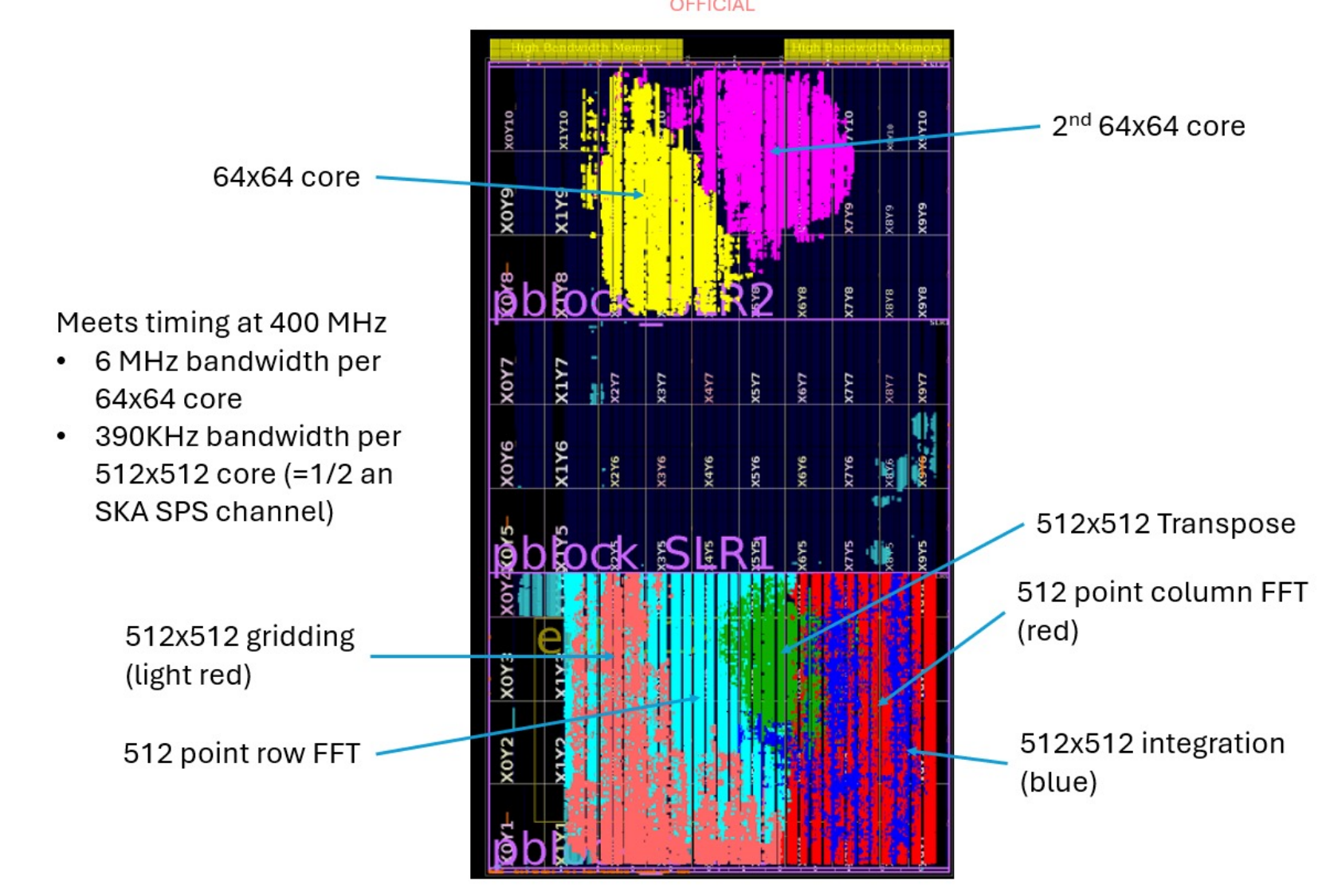}
   \end{tabular}
   \end{center}
   \caption[Resource utilisation on the \texttt{AMD Alveo V80} card] 
%>>>> use \label inside caption to get Fig. number with \ref{}
   { \label{fig:resources} Resource utilisation on the \texttt{AMD Alveo V80} card for two cores of 64$\times$64 processing and one core of 512$\times$512 processing.}
\end{figure}

Figure~\ref{fig:resources} shows the resource utilisation of the two prototypes on the \texttt{AMD Alveo V80} card. Each 64$\times$64 core can process $\sim 6$~MHz of dual-polarisation data, and thus the card has the capacity to process 8 such cores for a total bandwidth of $\sim 50$~MHz. Alternately, the card has the capacity to handle 780~kHz (one coarse channel) using the 512$\times$512 core with slightly different bitwidth optimisations. 

\section{Scalability and future outlook}\label{sec:future}

The demonstration of two different image sizes using the two prototype cores illustrates the scalability of the implementation of EPIC on the \texttt{AMD Alveo V80} FPGA card. It will be possible to scale to larger aperture arrays or the image size to larger number of pixels (increased field of view or angular resolution) by straightforwardly trading it against bandwidth and polarisation. The architecture is estimated to scale up to images of size $\sim 512\times 512$ or input aperture array sizes of few tens of thousands of elements before overflowing the card's memory. 

These preliminary results suggest that it outperforms the existing GPU implementation on LWA-Sevilleta by factor of several in bandwidth, and potentially at least an order of magnitude in array size, time resolution, and image size. The GPU implementations can improve in the future as well. However, an FPGA-based implementation of EPIC will remain a strong contender for large-$N$ aperture arrays of the future. 

\section{Summary}\label{sec:summary}

We are witnessing an advent of several large aperture arrays comprised of tens of thousands of elements with a primary goal of doing transformational science in the time domain. While aperture arrays open new parts of parameter space with their large fields of view and flexibility in operational modes, they also bring a formidable challenge of being extremely data intensive. This warrants a careful study and application of real-time processing algorithms and architectures. 

In an effort to address this challenge, we describe our preliminary exploration of a general purpose direct imaging architecture called E-field Parallel Imaging ``Correlator'' (EPIC), which is expected to hold significant advantage over traditional correlator and beamformer architectures in the case of large and dense aperture arrays. As FPGAs can be suited for highly efficient parallel implementation of low-bitwidth FFT, we are developing an FPGA-based prototype of EPIC and studying its performance towards scalability of aperture array and image sizes. Preliminary results indicate that our prototype can outperform current implementations of GPU-based EPIC and is scalable to arrays as large as few tens of thousands of elements and nearly 1~megapixel images with adequate optimisations. 
Further advances in this area will make EPIC and such FPGA implementations strong contenders for forming the backbone of real-time processing and for limiting compromises in discovery parameter space with large aperture arrays in the future. 

% \acknowledgments % equivalent to \section*{ACKNOWLEDGMENTS}       
 
% This unnumbered section is used to identify those who have aided the authors in understanding or accomplishing the work presented and to acknowledge sources of funding.  

% References
\bibliography{refs} % bibliography data in refs.bib

@ARTICLE{Bochenek+2020,
       author = {{Bochenek}, C.~D. and {Ravi}, V. and {Belov}, K.~V. and {Hallinan}, G. and {Kocz}, J. and {Kulkarni}, S.~R. and {McKenna}, D.~L.},
        title = "{A fast radio burst associated with a Galactic magnetar}",
      journal = {Nature},
         year = 2020,
        month = nov,
       volume = {587},
       number = {7832},
        pages = {59-62},
          doi = {10.1038/s41586-020-2872-x},
archivePrefix = {arXiv},
       eprint = {2005.10828},
 primaryClass = {astro-ph.HE},
       adsurl = {https://ui.adsabs.harvard.edu/abs/2020Natur.587...59B}
}

@ARTICLE{BURSTT,
       author = {{Lin}, Hsiu-Hsien and {Lin}, Kai-yang and {Li}, Chao-Te and {Tseng}, Yao-Huan and {Jiang}, Homin and {Wang}, Jen-Hung and {Cheng}, Jen-Chieh and {Pen}, Ue-Li and {Chen}, Ming-Tang and {Chen}, Pisin and {Chen}, Yaocheng and {Goto}, Tomotsugu and {Hashimoto}, Tetsuya and {Hwang}, Yuh-Jing and {King}, Sun-Kun and {Kubo}, Derek and {Kuo}, Chung-Yun and {Mills}, Adam and {Nam}, Jiwoo and {Oshiro}, Peter and {Shen}, Chang-Shao and {Tseng}, Hsien-Chun and {Wang}, Shih-Hao and {Wu}, Vigo Feng-Shun and {Bower}, Geoffrey and {Chang}, Shu-Hao and {Chen}, Pai-An and {Chen}, Ying-Chih and {Chiang}, Yi-Kuan and {Fedynitch}, Anatoli and {Gusinskaia}, Nina and {Ho}, Simon C.-C. and {Hsiao}, Tiger Y.-Y. and {Hu}, Chin-Ping and {Huang}, Yau De and {J{\'a}uregui Garc{\'\i}a}, Jos{\'e} Miguel and {Kim}, Seong Jin and {Kuo}, Cheng-Yu and {Ling}, Decmend Fang-Jie and {On}, Alvina Y.~L. and {Peterson}, Jeffrey B. and {R. Raquel}, Bjorn Jasper and {Su}, Shih-Chieh and {Uno}, Yuri and {Wu}, Cossas K.-W. and {Yamasaki}, Shotaro and {Zhu}, Hong-Ming},
        title = "{BURSTT: Bustling Universe Radio Survey Telescope in Taiwan}",
      journal = {PASP},
         year = 2022,
        month = sep,
       volume = {134},
       number = {1039},
          eid = {094106},
        pages = {094106},
          doi = {10.1088/1538-3873/ac8f71},
archivePrefix = {arXiv},
       eprint = {2206.08983},
 primaryClass = {astro-ph.IM},
       adsurl = {https://ui.adsabs.harvard.edu/abs/2022PASP..134i4106L}
}

@ARTICLE{Chandra+2016,
       author = {{Chandra}, Poonam and {Anupama}, G.~C. and {Arun}, K.~G. and {Iyyani}, Shabnam and {Misra}, Kuntal and {Narasimha}, D. and {Ray}, Alak and {Resmi}, L. and {Roy}, Subhashis and {Sutaria}, Firoza},
        title = "{Explosive and Radio-Selected Transients: Transient Astronomy with Square Kilometre Array and its Precursors}",
      journal = {Journal of Astrophysics and Astronomy},
         year = 2016,
        month = dec,
       volume = {37},
       number = {4},
          eid = {30},
        pages = {30},
          doi = {10.1007/s12036-016-9408-7},
archivePrefix = {arXiv},
       eprint = {1610.08178},
 primaryClass = {astro-ph.HE},
       adsurl = {https://ui.adsabs.harvard.edu/abs/2016JApA...37...30C}
}

@INPROCEEDINGS{Daishido+1991,
       author = {{Daishido}, Tsuneaki and {Asuma}, Kuniyuki and {Nishibori}, Kazuhiko and {Nakajima}, Junichi and {Yano}, Motoko and {Otobe}, Eiichiro and {Watanabe}, Naoki and {Tsuchiya}, Akira and {Iwase}, Seiichiro},
        title = "{Direct imaging digital lens for transient radio source survey}",
    booktitle = {IAU Colloq. 131: Radio Interferometry. Theory, Techniques, and Applications},
         year = 1991,
       editor = {{Cornwell}, T.~J. and {Perley}, R.~A.},
       series = {Astronomical Society of the Pacific Conference Series},
       volume = {19},
        month = jan,
        pages = {86-89},
       adsurl = {https://ui.adsabs.harvard.edu/abs/1991ASPC...19...86D}
}

@ARTICLE{Dewdney+2009,
       author = {{Dewdney}, P.~E. and {Hall}, P.~J. and {Schilizzi}, R.~T. and {Lazio}, T.~J.~L.~W.},
        title = "{The Square Kilometre Array}",
      journal = {IEEE Proceedings},
         year = 2009,
        month = aug,
       volume = {97},
       number = {8},
        pages = {1482-1496},
          doi = {10.1109/JPROC.2009.2021005},
       adsurl = {https://ui.adsabs.harvard.edu/abs/2009IEEEP..97.1482D}
}

@ARTICLE{Dowell+2018,
       author = {{Dowell}, Jayce and {Taylor}, Greg B.},
        title = "{The Swarm Telescope Concept}",
      journal = {Journal of Astronomical Instrumentation},
         year = 2018,
        month = jan,
       volume = {7},
          eid = {1850006},
        pages = {1850006},
          doi = {10.1142/S225117171850006X},
archivePrefix = {arXiv},
       eprint = {1806.10634},
 primaryClass = {astro-ph.IM},
       adsurl = {https://ui.adsabs.harvard.edu/abs/2018JAI.....750006D}
}

@INPROCEEDINGS{DSA-2000,
       author = {{Hallinan}, Gregg and {Ravi}, V. and {Weinreb}, S. and {Kocz}, J. and {Huang}, Y. and {Woody}, D.~P. and {Lamb}, J. and {D'Addario}, L. and {Catha}, M. and {Law}, C. and {Kulkarni}, S.~R. and {Phinney}, E.~S. and {Eastwood}, M.~W. and {Bouman}, K. and {McLaughlin}, M. and {Ransom}, S. and {Siemens}, X. and {Cordes}, J. and {Lynch}, R. and {Kaplan}, D. and {Brazier}, A. and {Bhatnagar}, S. and {Myers}, S. and {Walter}, F. and {Gaensler}, B.},
        title = "{The DSA-2000 {\textemdash} A Radio Survey Camera}",
    booktitle = {Bulletin of the American Astronomical Society},
         year = 2019,
       volume = {51},
        month = sep,
          eid = {255},
        pages = {255},
          doi = {10.48550/arXiv.1907.07648},
archivePrefix = {arXiv},
       eprint = {1907.07648},
 primaryClass = {astro-ph.IM},
       adsurl = {https://ui.adsabs.harvard.edu/abs/2019BAAS...51g.255H}
}

@ARTICLE{Foster+2014,
       author = {{Foster}, G. and {Hickish}, J. and {Magro}, A. and {Price}, D. and {Zarb Adami}, K.},
        title = "{Implementation of a direct-imaging and FX correlator for the BEST-2 array}",
      journal = {MNRAS},
         year = 2014,
        month = apr,
       volume = {439},
       number = {3},
        pages = {3180-3188},
          doi = {10.1093/mnras/stu188},
archivePrefix = {arXiv},
       eprint = {1401.6753},
 primaryClass = {astro-ph.IM},
       adsurl = {https://ui.adsabs.harvard.edu/abs/2014MNRAS.439.3180F}
}

@ARTICLE{Hampson+2025, 
    title={SKA LOW Correlator and Beamformer Signal Processing}, 
    url={http://dx.doi.org/10.46620/ursiaprasc25/icar1244}, 
    DOI={10.46620/ursiaprasc25/icar1244}, 
    journal={Proceedings of the 7th URSI Asia-Pacific RadioScience Conference – AP-RASC 2025}, 
    publisher={URSI – International Union of Radio Science}, 
    author={Hampson, Grant and Bunton, John and Chen, Yuqing and Humphrey, David and Babich, Giles and Bengston, Keith and Bolin, Andrew and Jourjon, Guillaume and Bacic, Bernardo}, year={2025} 
}

@ARTICLE{HIRAX+2022,
       author = {{Crichton}, Devin and {Aich}, Moumita and {Amara}, Adam and {Bandura}, Kevin and {Bassett}, Bruce A. and {Bengaly}, Carlos and {Berner}, Pascale and {Bhatporia}, Shruti and {Bucher}, Martin and {Chang}, Tzu-Ching and {Chiang}, H. Cynthia and {Cliche}, Jean-Francois and {Crichton}, Carolyn and {Dave}, Romeel and {De Villiers}, Dirk I.~L. and {Dobbs}, Matt and {Ewall-Wice}, Aaron M. and {Eyono}, Scott and {Finlay}, Christopher and {Gaddam}, Sindhu and {Ganga}, Ken and {Gayley}, Kevin G. and {Gerodias}, Kit and {Gibbon}, Tim B. and {Gumba}, Austine and {Gupta}, Neeraj and {Harris}, Maile and {Heilgendorff}, Heiko and {Hilton}, Matt and {Hincks}, Adam D. and {Hitz}, Pascal and {Jalilvand}, Mona and {Julie}, Roufurd P.~M. and {Kader}, Zahra and {Kania}, Joseph and {Karagiannis}, Dionysios and {Karastergiou}, Aris and {Kesebonye}, Kabelo and {Kittiwisit}, Piyanat and {Kneib}, Jean-Paul and {Knowles}, Kenda and {Kuhn}, Emily R. and {Kunz}, Martin and {Maartens}, Roy and {MacKay}, Vincent and {MacPherson}, Stuart and {Monstein}, Christian and {Moodley}, Kavilan and {Mugundhan}, V. and {Naidoo}, Warren and {Naidu}, Arun and {Newburgh}, Laura B. and {Nistane}, Viraj and {Di Nitto}, Amanda and {{\"O}l{\c{c}}ek}, Deniz and {Pan}, Xinyu and {Paul}, Sourabh and {Peterson}, Jeffrey B. and {Pieters}, Elizabeth and {Pieterse}, Carla and {Pillay}, Aritha and {Polish}, Anna R. and {Randrianjanahary}, Liantsoa and {Refregier}, Alexandre and {Renard}, Andre and {Retana-Montenegro}, Edwin and {Rout}, Ian H. and {Russeeawon}, Cyndie and {Sadr}, Alireza Vafaei and {Saliwanchik}, Benjamin R.~B. and {Sampath}, Ajith and {Sanghavi}, Pranav and {Santos}, Mario G. and {Sengate}, Onkabetse and {Shaw}, J. Richard and {Sievers}, Jonathan L. and {Smirnov}, Oleg M. and {Smith}, Kendrick M. and {Sob}, Ulrich Armel Mbou and {Srianand}, Raghunathan and {Stronkhorst}, Pieter and {Sunder}, Dhaneshwar D. and {Tartakovsky}, Simon and {Taylor}, Russ and {Timbie}, Peter and {Tolley}, Emma E. and {Townsend}, Junaid and {Tyndall}, Will and {Ungerer}, Cornelius and {van Dyk}, Jacques and {van Vuuren}, Gary and {Vanderlinde}, Keith and {Viant}, Thierry and {Walters}, Anthony and {Wang}, Jingying and {Weltman}, Amanda and {Woudt}, Patrick and {Wulf}, Dallas and {Zavyalov}, Anatoly and {Zhang}, Zheng},
        title = "{Hydrogen Intensity and Real-Time Analysis Experiment: 256-element array status and overview}",
      journal = {Journal of Astronomical Telescopes, Instruments, and Systems},
         year = 2022,
        month = jan,
       volume = {8},
          eid = {011019},
        pages = {011019},
          doi = {10.1117/1.JATIS.8.1.011019},
archivePrefix = {arXiv},
       eprint = {2109.13755},
 primaryClass = {astro-ph.IM},
       adsurl = {https://ui.adsabs.harvard.edu/abs/2022JATIS...8a1019C}
}

@INPROCEEDINGS{Keane2013,
       author = {{Keane}, E.~F.},
        title = "{Radio pulsar variability}",
    booktitle = {Neutron Stars and Pulsars: Challenges and Opportunities after 80 years},
         year = 2013,
       editor = {{van Leeuwen}, Joeri},
       volume = {291},
        month = mar,
        pages = {295-300},
          doi = {10.1017/S1743921312023927},
archivePrefix = {arXiv},
       eprint = {1210.5397},
 primaryClass = {astro-ph.SR},
       adsurl = {https://ui.adsabs.harvard.edu/abs/2013IAUS..291..295K}
}

@ARTICLE{Kent+2019,
       author = {{Kent}, James and {Dowell}, Jayce and {Beardsley}, Adam and {Thyagarajan}, Nithyanandan and {Taylor}, Greg and {Bowman}, Judd},
        title = "{A real-time, all-sky, high time resolution, direct imager for the long wavelength array}",
      journal = {MNRAS},
         year = 2019,
        month = jul,
       volume = {486},
       number = {4},
        pages = {5052-5060},
          doi = {10.1093/mnras/stz1206},
archivePrefix = {arXiv},
       eprint = {1904.11422},
 primaryClass = {astro-ph.IM},
       adsurl = {https://ui.adsabs.harvard.edu/abs/2019MNRAS.486.5052K}
}

@ARTICLE{Kent+2020,
       author = {{Kent}, James and {Beardsley}, Adam P. and {Bester}, Landman and {Gull}, Steve F. and {Nikolic}, Bojan and {Dowell}, Jayce and {Thyagarajan}, Nithyanandan and {Taylor}, Greg B. and {Bowman}, Judd},
        title = "{Direct wide-field radio imaging in real-time at high time resolution using antenna electric fields}",
      journal = {MNRAS},
         year = 2020,
        month = jan,
       volume = {491},
       number = {1},
        pages = {254-263},
          doi = {10.1093/mnras/stz3028},
archivePrefix = {arXiv},
       eprint = {1909.03973},
 primaryClass = {astro-ph.IM},
       adsurl = {https://ui.adsabs.harvard.edu/abs/2020MNRAS.491..254K}
}

@ARTICLE{Krishnan+2023,
       author = {{Krishnan}, Harihanan and {Beardsley}, Adam P. and {Bowman}, Judd D. and {Dowell}, Jayce and {Kolopanis}, Matthew and {Taylor}, Greg and {Thyagarajan}, Nithyanandan},
        title = "{Optimization and commissioning of the EPIC commensal radio transient imager for the long wavelength array}",
      journal = {MNRAS},
         year = 2023,
        month = apr,
       volume = {520},
       number = {2},
        pages = {1928-1937},
          doi = {10.1093/mnras/stad263},
archivePrefix = {arXiv},
       eprint = {2301.09662},
 primaryClass = {astro-ph.IM},
       adsurl = {https://ui.adsabs.harvard.edu/abs/2023MNRAS.520.1928K}
}

@ARTICLE{Lorimer+2007,
       author = {{Lorimer}, D.~R. and {Bailes}, M. and {McLaughlin}, M.~A. and {Narkevic}, D.~J. and {Crawford}, F.},
        title = "{A Bright Millisecond Radio Burst of Extragalactic Origin}",
      journal = {Science},
         year = 2007,
        month = nov,
       volume = {318},
       number = {5851},
        pages = {777},
          doi = {10.1126/science.1147532},
archivePrefix = {arXiv},
       eprint = {0709.4301},
 primaryClass = {astro-ph},
       adsurl = {https://ui.adsabs.harvard.edu/abs/2007Sci...318..777L}
}

@ARTICLE{Macario+2022,
       author = {{Macario}, Giulia and {Pupillo}, Giuseppe and {Bernardi}, Gianni and {Bolli}, Pietro and {Di Ninni}, Paola and {Comoretto}, Giovanni and {Mattana}, Andrea and {Monari}, Jader and {Perini}, Federico and {Schiaffino}, Marco and {Sokolowski}, Marcin and {Wayth}, Randall and {Broderick}, Jess and {Waterson}, Mark and {Grazia Labate}, Maria and {Chiello}, Riccardo and {Magro}, Alessio and {Booler}, Tom and {Mcphail}, Andrew and {Minchin}, Dave and {Bhushan}, Raunaq},
        title = "{Characterization of the SKA1-Low prototype station Aperture Array Verification System 2}",
      journal = {Journal of Astronomical Telescopes, Instruments, and Systems},
         year = 2022,
        month = jan,
       volume = {8},
          eid = {011014},
        pages = {011014},
          doi = {10.1117/1.JATIS.8.1.011014},
archivePrefix = {arXiv},
       eprint = {2109.11983},
 primaryClass = {astro-ph.IM},
       adsurl = {https://ui.adsabs.harvard.edu/abs/2022JATIS...8a1014M}
}

@ARTICLE{Mclaughlin+2006,
       author = {{McLaughlin}, M.~A. and {Lyne}, A.~G. and {Lorimer}, D.~R. and {Kramer}, M. and {Faulkner}, A.~J. and {Manchester}, R.~N. and {Cordes}, J.~M. and {Camilo}, F. and {Possenti}, A. and {Stairs}, I.~H. and {Hobbs}, G. and {D'Amico}, N. and {Burgay}, M. and {O'Brien}, J.~T.},
        title = "{Transient radio bursts from rotating neutron stars}",
      journal = {Nature},
         year = 2006,
        month = feb,
       volume = {439},
       number = {7078},
        pages = {817-820},
          doi = {10.1038/nature04440},
archivePrefix = {arXiv},
       eprint = {astro-ph/0511587},
 primaryClass = {astro-ph},
       adsurl = {https://ui.adsabs.harvard.edu/abs/2006Natur.439..817M}
}

@ARTICLE{Murphy+Kaplan2026,
       author = {{Murphy}, Tara and {Kaplan}, David L.},
        title = "{The Dawes review 13: A new look at the dynamic radio sky}",
      journal = {PASA},
         year = 2026,
        month = jan,
       volume = {43},
          eid = {e006},
        pages = {e006},
          doi = {10.1017/pasa.2025.10128},
archivePrefix = {arXiv},
       eprint = {2511.10785},
 primaryClass = {astro-ph.SR},
       adsurl = {https://ui.adsabs.harvard.edu/abs/2026PASA...43....6M}
}

@ARTICLE{Nimmo+2022,
       author = {{Nimmo}, K. and {Hessels}, J.~W.~T. and {Kirsten}, F. and {Keimpema}, A. and {Cordes}, J.~M. and {Snelders}, M.~P. and {Hewitt}, D.~M. and {Karuppusamy}, R. and {Archibald}, A.~M. and {Bezrukovs}, V. and {Bhardwaj}, M. and {Blaauw}, R. and {Buttaccio}, S.~T. and {Cassanelli}, T. and {Conway}, J.~E. and {Corongiu}, A. and {Feiler}, R. and {Fonseca}, E. and {Forss{\'e}n}, O. and {Gawro{\'n}ski}, M. and {Giroletti}, M. and {Kharinov}, M.~A. and {Leung}, C. and {Lindqvist}, M. and {Maccaferri}, G. and {Marcote}, B. and {Masui}, K.~W. and {Mckinven}, R. and {Melnikov}, A. and {Michilli}, D. and {Mikhailov}, A.~G. and {Ng}, C. and {Orbidans}, A. and {Ould-Boukattine}, O.~S. and {Paragi}, Z. and {Pearlman}, A.~B. and {Petroff}, E. and {Rahman}, M. and {Scholz}, P. and {Shin}, K. and {Smith}, K.~M. and {Stairs}, I.~H. and {Surcis}, G. and {Tendulkar}, S.~P. and {Vlemmings}, W. and {Wang}, N. and {Yang}, J. and {Yuan}, J.~P.},
        title = "{Burst timescales and luminosities as links between young pulsars and fast radio bursts}",
      journal = {Nature Astronomy},
         year = 2022,
        month = feb,
       volume = {6},
        pages = {393-401},
          doi = {10.1038/s41550-021-01569-9},
archivePrefix = {arXiv},
       eprint = {2105.11446},
 primaryClass = {astro-ph.HE},
       adsurl = {https://ui.adsabs.harvard.edu/abs/2022NatAs...6..393N}
}

@ARTICLE{Otobe+1994,
       author = {{Otobe}, Eiichiro and {Nakajima}, Junichi and {Nishibori}, Kazuhiko and {Saito}, Tomohiro and {Kobayashi}, Hiromi and {Tanaka}, Naoki and {Watanabe}, Naoki and {Aramaki}, Yoshitaka and {Hoshikawa}, Tomoyuki and {Asuma}, Kuniyuki and {Daishido}, Tsuneaki},
        title = "{Two-Dimensional Direct Images with a Spatial FFT Interferometer}",
      journal = {PASJ},
         year = 1994,
        month = oct,
       volume = {46},
        pages = {503-510},
       adsurl = {https://ui.adsabs.harvard.edu/abs/1994PASJ...46..503O}
}

@ARTICLE{Pietka+2015,
       author = {{Pietka}, M. and {Fender}, R.~P. and {Keane}, E.~F.},
        title = "{The variability time-scales and brightness temperatures of radio flares from stars to supermassive black holes}",
      journal = {MNRAS},
         year = 2015,
        month = feb,
       volume = {446},
       number = {4},
        pages = {3687-3696},
          doi = {10.1093/mnras/stu2335},
archivePrefix = {arXiv},
       eprint = {1411.1067},
 primaryClass = {astro-ph.HE},
       adsurl = {https://ui.adsabs.harvard.edu/abs/2015MNRAS.446.3687P}
}

@ARTICLE{Price2024,
       author = {{Price}, D.~C.},
        title = "{Reduced-resolution beamforming: Lowering the computational cost for pulsar and technosignature surveys}",
      journal = {PASA},
         year = 2024,
        month = may,
       volume = {41},
          eid = {e037},
        pages = {e037},
          doi = {10.1017/pasa.2024.35},
archivePrefix = {arXiv},
       eprint = {2402.12723},
 primaryClass = {astro-ph.IM},
       adsurl = {https://ui.adsabs.harvard.edu/abs/2024PASA...41...37P}
}

@ARTICLE{Rea+2026,
       author = {{Rea}, Nanda and {Hurley-Walker}, Natasha and {Caleb}, Manisha},
        title = "{Long period transients (LPTs): A comprehensive review}",
      journal = {Journal of High Energy Astrophysics},
         year = 2026,
        month = apr,
       volume = {52},
          eid = {100566},
        pages = {100566},
          doi = {10.1016/j.jheap.2026.100566},
archivePrefix = {arXiv},
       eprint = {2601.10393},
 primaryClass = {astro-ph.HE},
       adsurl = {https://ui.adsabs.harvard.edu/abs/2026JHEAp..5200566R}
}

@inproceedings{Reddy+2024,
author = {Karthik Reddy and Judd D. Bowman and Jayce Dowell and Greg B. Taylor and Adam P. Beardsley and Craig Taylor},
title = {{Architecture and implementation of a 25000 FPS radio camera on the long wavelength array}},
volume = {13101},
booktitle = {Software and Cyberinfrastructure for Astronomy VIII},
editor = {Jorge Ibsen and Gianluca Chiozzi},
organization = {International Society for Optics and Photonics},
publisher = {SPIE},
pages = {131011C},
year = {2024},
doi = {10.1117/12.3012524},
URL = {https://doi.org/10.1117/12.3012524}
}

@techreport{LWA-EPIC-memo9,
  author      = {Karthik Reddy},
  title       = {EPIC code optimizations},
  institution = {Arizona State University},
  url         = {https://github.com/epic-astronomy/Memos/blob/master/PDFs/009_EPIC_Code_Optimizations.md}, 
  type        = {EPIC Memorandum},
  number      = {9},
  year        = {2023},
  note        = {Available at \url{https://github.com/epic-astronomy/Memos/blob/master/PDFs/009_EPIC_Code_Optimizations.md}}
}

@ARTICLE{SKA1+2019,
       author = {{Braun}, Robert and {Bonaldi}, Anna and {Bourke}, Tyler and {Keane}, Evan and {Wagg}, Jeff},
        title = "{Anticipated Performance of the Square Kilometre Array -- Phase 1 (SKA1)}",
      journal = {arXiv e-prints},
         year = 2019,
        month = dec,
          eid = {arXiv:1912.12699},
        pages = {arXiv:1912.12699},
          doi = {10.48550/arXiv.1912.12699},
archivePrefix = {arXiv},
       eprint = {1912.12699},
 primaryClass = {astro-ph.IM},
       adsurl = {https://ui.adsabs.harvard.edu/abs/2019arXiv191212699B}
}

@ARTICLE{Thornton+2013,
       author = {{Thornton}, D. and {Stappers}, B. and {Bailes}, M. and {Barsdell}, B. and {Bates}, S. and {Bhat}, N.~D.~R. and {Burgay}, M. and {Burke-Spolaor}, S. and {Champion}, D.~J. and {Coster}, P. and {D'Amico}, N. and {Jameson}, A. and {Johnston}, S. and {Keith}, M. and {Kramer}, M. and {Levin}, L. and {Milia}, S. and {Ng}, C. and {Possenti}, A. and {van Straten}, W.},
        title = "{A Population of Fast Radio Bursts at Cosmological Distances}",
      journal = {Science},
         year = 2013,
        month = jul,
       volume = {341},
       number = {6141},
        pages = {53-56},
          doi = {10.1126/science.1236789},
archivePrefix = {arXiv},
       eprint = {1307.1628},
 primaryClass = {astro-ph.HE},
       adsurl = {https://ui.adsabs.harvard.edu/abs/2013Sci...341...53T}
}

@ARTICLE{Thyagarajan+2017,
       author = {{Thyagarajan}, Nithyanandan and {Beardsley}, Adam P. and {Bowman}, Judd D. and {Morales}, Miguel F.},
        title = "{A Generic and Efficient E-field Parallel Imaging Correlator for Next-Generation Radio Telescopes}",
      journal = {MNRAS},
         year = 2017,
        month = may,
       volume = {467},
       number = {1},
        pages = {715-730},
          doi = {10.1093/mnras/stx113},
archivePrefix = {arXiv},
       eprint = {1510.08318},
 primaryClass = {astro-ph.IM},
       adsurl = {https://ui.adsabs.harvard.edu/abs/2017MNRAS.467..715T}
}
\bibliographystyle{spiebib} % makes bibtex use spiebib.bst

\end{document}